\documentclass[twocolumn]{aastex701}
\usepackage{url}

\begin{document}
%
\title{Gamma-ray Quasi-periodic Oscillations in the Changing Look Blazar B2 1420+326 (OQ 334)} 
\author[orcid=0000-0001-8282-4561]{Sunanda}
\affiliation{Aryabhatta Research Institute of Observational Sciences (ARIES), Manora Peak, Nainital 263001, India}
\email[show]{ss7010594@gmail.com}
\author[orcid=0000-0001-8716-9412]{Shubham Kishore} 
\affiliation{Indian Institute of Astrophysics, 2nd Block, Koramangala, Bangalore 560034, India}
\email[show]{amp700151@gmail.com}
\author[orcid=0000-0002-9331-4388]{Alok C. Gupta}
\affiliation{Aryabhatta Research Institute of Observational Sciences (ARIES), Manora Peak, Nainital 263001, India}
\email[]{acgupta30@gmail.com}  
\begin{abstract}
In the \textit{Fermi}-LAT gamma-ray light curve from January 1 to August 25, 2026, of the changing look blazar B2 1420+326, we report here the detection of transient quasi-periodic oscillations (QPOs) of period ranging around 10 – 13 d.
The source entered the flare state around May 7, 2026, and lasted till around mid-August,  2026 during which it depicted strong QPO signatures. We utilized standard techniques, including weighted wavelet Z-transform analysis, generalized Lomb–Scargle periodogram, and phase dispersion minimization, to search for and confirm these findings. Prior to the flare, the source displayed a mild damped oscillatory feature of similar periodicity with a damping timescale of around 20~d. Based on the period timescale and nature of the QPOs detected, we discuss possible physical mechanisms for the observed salient features.
\end{abstract}
\section{Introduction}
\noindent
Active galactic nuclei (AGNs) apparently have accreting supermassive black holes (SMBHs) with masses of 10$^{6} – \rm{10}^{10} \ \rm{M}_{\odot}$ and have many similarities to scaled-up galactic X-ray emitting black hole (BH) binaries. Despite the fact that extensive searches of quasi-periodic oscillations (QPOs) in AGNs light curves (LCs) have been done, but they are rarely found \citep[e.g.][and references therein]{2008Natur.455..369G,2022Natur.609..265J,2025MNRAS.541.2955P}, although they are very often detected in BH and neutron star binaries in our and nearby galaxies \citep[e.g.][]{2006ARA&A..44...49R}. Blazar is a subclass of AGN that emits radiation in the whole electromagnetic (EM) spectrum, and emission is predominantly nonthermal. Blazars are further classified into BL Lacertae objects (BL Lacs) and flat spectrum radio quasars (FSRQs). In the composite optical/UV (ultraviolet) spectrum, BL Lacs show a featureless spectrum (i.e., the equivalent width (EW) of emission line is $\leq \rm{5}\AA$), whereas FSRQs show prominent emission lines. Blazars show variable flux, spectra, and polarization in the complete EM spectrum on diverse timescales ranging as short as a few minutes to as long as several years. \\
\\
The search for QPOs in different subclasses of AGN LCs has become a significant scientific project in extragalactic astronomy, utilizing a variety of analysis techniques used in astronomy and in the stock market e.g., weighted wavelet Z analysis (WWZ), Lomb–Scargle periodogram (LSP), power spectral density (PSD), phase dispersion minimization (PDM), autoregressive (AR) integrated (I) moving average (MA) or ARIMA(p, d, q) model, etc. \citep[e.g.][and references therein]{2021MNRAS.501...50S,2026MNRAS.550g1284K}, following the discovery of strong evidence of a QPO with a period of $\sim$1 hour in the X-ray LC of the AGN RE J1034+396 \citep{2008Natur.455..369G}. There have been occasional strong claims of QPO detection in several blazars on diverse timescales in a single EM band, transient QPOs in a single EM band, and also detection of QPOs simultaneously in more than one EM band \citep[e.g.,][and references therein]{2009ApJ...690..216G,2015ApJ...813L..41A,2015Natur.518...74G,2018NatCo...9.4599Z,2022Natur.609..265J,2025MNRAS.541.2955P,2026MNRAS.550g1284K}. There have also been a few occasional strong claims of QPO detection in other classes of AGNs in a single EM band \citep[e.g.,][and references therein]{2008Natur.455..369G,2018A&A...616L...6G,2025ApJ...992L..13Z}. \\
\\
The blazar B2 1420+326 (OQ 334, TXS 1420+326, PKS 1420+326) at a redshift of z = 0.6819 \citep{2010MNRAS.405.2302H} has been classified as FSRQ in the 4FGL catalog \citep{2020ApJS..247...33A}. The source entered a high flux state in the year 2017, and it underwent multiple episodes of $\gamma-$ray flaring till 2020, and the detailed optical spectroscopic monitoring of the source revealed its transitional spectral feature between FSRQ and BL Lac, and vice versa, and classified it as a changing-look (CL) blazar \citep{2021ApJ...913..146M}. CL AGNs are those where nuclear processes like accretion and obscuration cause emission lines to emerge and vanish. Emission lines typically show up in these sources when the source is brighter. Over a coarse time span (a few year's duration), they exhibit varying emission line strength. A few hundred such CL AGNs have been detected and studied on multiple occasions in the last about two decades \citep[e.g.][and references therein]{2005A&A...442..185B,2016MNRAS.457..389M,2019ApJ...874....8M,2026arXiv260715564Z}. In this case, it is important to distinguish between CL  blazars and CL AGNs because in blazars, the emission lines vanish as the source brightens due to an entirely different mechanism—that is, a rise in the featureless jet continuum. FSRQs frequently resemble BL Lacs during outbursts, vice versa \citep[e.g.][and references therein]{2017Natur.552..374R,2025ApJ...978..120P}. CL blazars are much rarer in comparison to AGNs \citep[e.g.][and references therein]{2017Natur.552..374R,2021ApJ...913..146M,2024ApJ...962..122K,2025A&A...703L..13A,2025ApJ...978..120P,2026MNRAS.551g1356R}, and to the best of our knowledge, only four such blazars: AO 0235+164 \citep{2026MNRAS.551g1356R}, J1243+4043 \citep{2025A&A...703L..13A}, B2 1308+326 \citep{2025ApJ...978..120P}, and B2 1420+326 \citep{2021ApJ...913..146M} have been studied in more detail. \\
\\
In this study, we report the $\gamma$-ray QPO detection with the period of $\sim$10 d in the CL blazar B2 1420+326, which is the first result of QPO detection from this source in the $\gamma$-ray band. For this study, we have analyzed \textit{Fermi}-LAT $\gamma-$ray data of the source taken in the year 2026. Section \ref{section2} details the \textit{Fermi}-LAT data acquisition and analysis; data analysis is provided in Section \ref{section3}. The results are reported in Section \ref{results}, followed by a discussion and conclusions in Section \ref{section5}.

\section{Data Acquisition}
\label{section2}
\subsection{Fermi-LAT data}
\noindent
The high-energy $\gamma$-rays data for B2 1420+326 were obtained from the Large Area Telescope (LAT) aboard the \textit{Fermi} Space Telescope. \textit{Fermi}-LAT is a space-based imaging telescope that uses the pair-production approach to observe the $\gamma$-rays over the energy range  $30~\mathrm{MeV} - 1~\mathrm{TeV}$. It has a wide angular field of view of approximately 2.3 sr and surveys the whole sky roughly every three hours \citep{2009ApJ...697.1071A}. For B2 1420+326, we retrieved the PASS 8 event data during the  MJD 61041 to MJD 61277 (January 1, 2026 to  August 25, 2026) from the \textit{Fermi}-LAT repository\footnote{\url{https://fermi.gsfc.nasa.gov/cgi-bin/ssc/LAT/LATDataQuery.cgi}}.
\subsection{Data Reduction}
\begin{figure*}
    \includegraphics[width=1\linewidth]{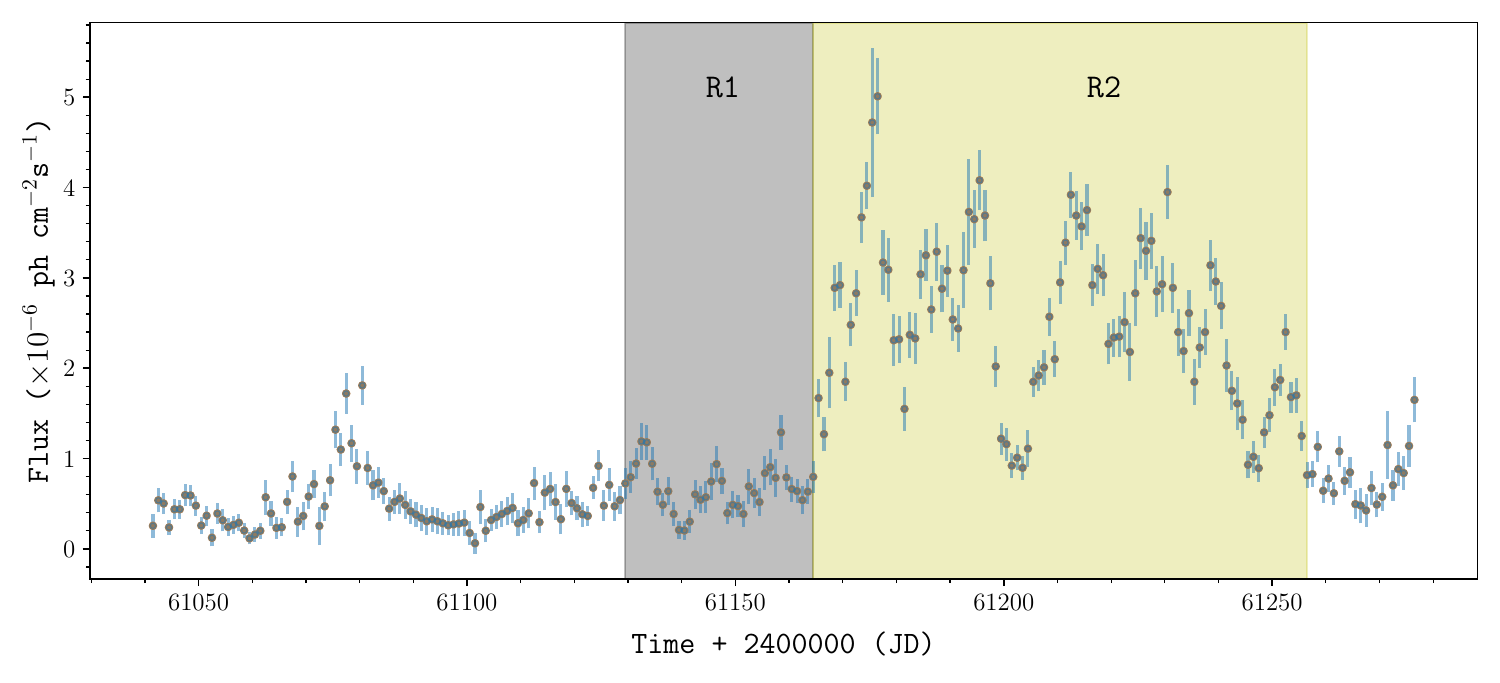}
    \caption{$\gamma$-ray light curve spanning Jan. 01, 2026 to Aug, 25, 2026. The two shaded regions, R1 and R2, highlight the regions of interest. R2 marks the flare spans, selected visually.}
    \label{fig_1}
\end{figure*}

\noindent
The standard software FERMITOOLS-V2.2.0, together with the user-developed Python script ENRICO \citep{2013ICRC...33.2784S}, was used for unbinned maximum-likelihood analysis. The \textit{Fermi}-LAT data of the source were obtained within the region of interest (ROI) $15^\circ$ centered at the source in the energy range $0.1\text{--}300~\mathrm{GeV}$.  To avoid contamination of the $\gamma$-ray data from the Earth's limb, we applied a zenith-angle cut of less than $90^\circ$.  For the good time intervals (GTIs), we used a filter \texttt{DATA\_QUAL > 0 \&\& LAT\_CONFIG == 1}. We selected the events belonging to the Source class $(\mathrm{evclass}=128)$ and included both front+ back event types ($\mathrm{evtype}=3$). The instrumental response function (IRF) \text{``${P8R3}{\_}{SOURCE}{\_}{V3}$''} ,  isotropic background model \text{``$\textup{{iso}{\_}{P8R3}{\_}{SOURCE}{\_}{V3}{\_}{v06}.{txt}}$''}, and Galactic diffused model \text{``$\textup{{gll}{\_}{iem}{\_}v07.fits}$''} were employed for the analysis. The model file included all sources from the Fourth Fermi-LAT Source Catalog (4FGL) located within ROI+10$^\circ$ of B2 1420+326.  Since each analysis was performed over a 24-hour interval, we used a simple power-law model for the spectrum of B2 1420+326 and included EBL correction from \cite{2008A&A...487..837F}. All spectral parameters associated with sources located within \(5^\circ\) of B2 1420+326 were kept free. The flux data points in the LC have test statistics (TS) $\ge$ 9, where  TS $=$9 corresponds approximately to  \(3\sigma\). For time bins with \(TS < 9\), 95$\%$ confidence-level upper limits were estimated. The obtained one-day binned LC in the energy range 0.1–300 GeV is shown in the Fig.~\ref{fig_1}.
%
\section{Data Analysis}
\label{section3}
\noindent
Here, we have acquired the \textit{Fermi} $\gamma$-ray  LC data (see Fig.~\ref{fig_1}) from January 1, 2026, to August 25, 2026. Visual inspection of the LC reveals a diverse nature of B2~1420+326, including both quiescent and variable (with flaring)  phases within a timespan of around eight months. Although the source entered a large flaring state around MJD~61167 during (May 7, 2026), and our interest mainly focuses on the sharply visible oscillatory features during this flare (marked as shaded region R2 in Fig.~\ref{fig_1}), another mildly damped oscillatory feature (marked as shaded region R1 in Fig.~\ref{fig_1}) can also be observed during an otherwise quiescent phase that lasted around a month just before entering the large flare. Additionally, the source exhibits a short flare in February 2026 (around MJD~61080). The scope of this paper lies in assessing the possible QPO features in different portions of the composite  LC of the source, and similar to \cite{2026MNRAS.550g1284K}, we have utilized and implemented the weighted wavelet Z \citep[WWZ;][]{1996AJ....112.1709F}, generalized Lomb-Scargle periodogram \citep[LSP; e.g.,][]{1976Ap&SS..39..447L, 1982ApJ...263..835S, 1986ApJ...302..757H, 2009A&A...496..577Z, 2018ApJS..236...16V}, and phase dispersion minimization \citep[PDM;][]{1978ApJ...224..953S} methods to test and deduce the conforming results. We request the readers, therefore, to hover over \citep{2026MNRAS.550g1284K} to get a collective overview of these methods employed.
\begin{figure}
    \centering
    \includegraphics[width=1\linewidth]{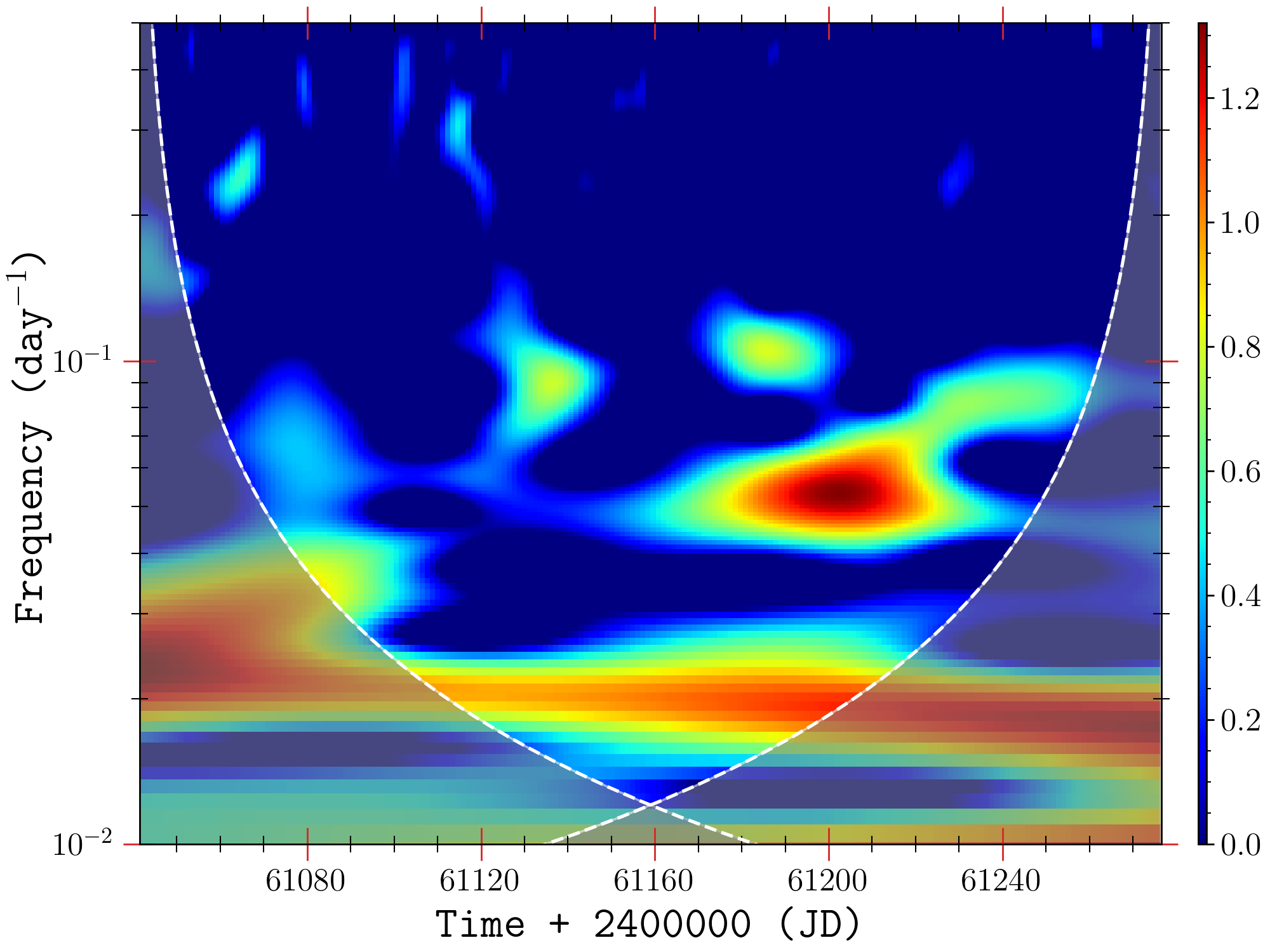}
    \caption{WWZ map of the complete $\gamma$-ray light curve; mildly shaded region under the white dashed parabolic type lines represents the cone of influence due to edge effects, where any QPO feature is unreliable. The power depicted is on a log scale for better visualization of the three salient QPO features at around $10^{-1}~\text{d}^{-1}$.}
    \label{fig_2}
\end{figure}
\section{Results}
\label{results}
\begin{figure}
    \centering
    \includegraphics[width=0.9\linewidth, trim=0 1cm 0 0, clip]{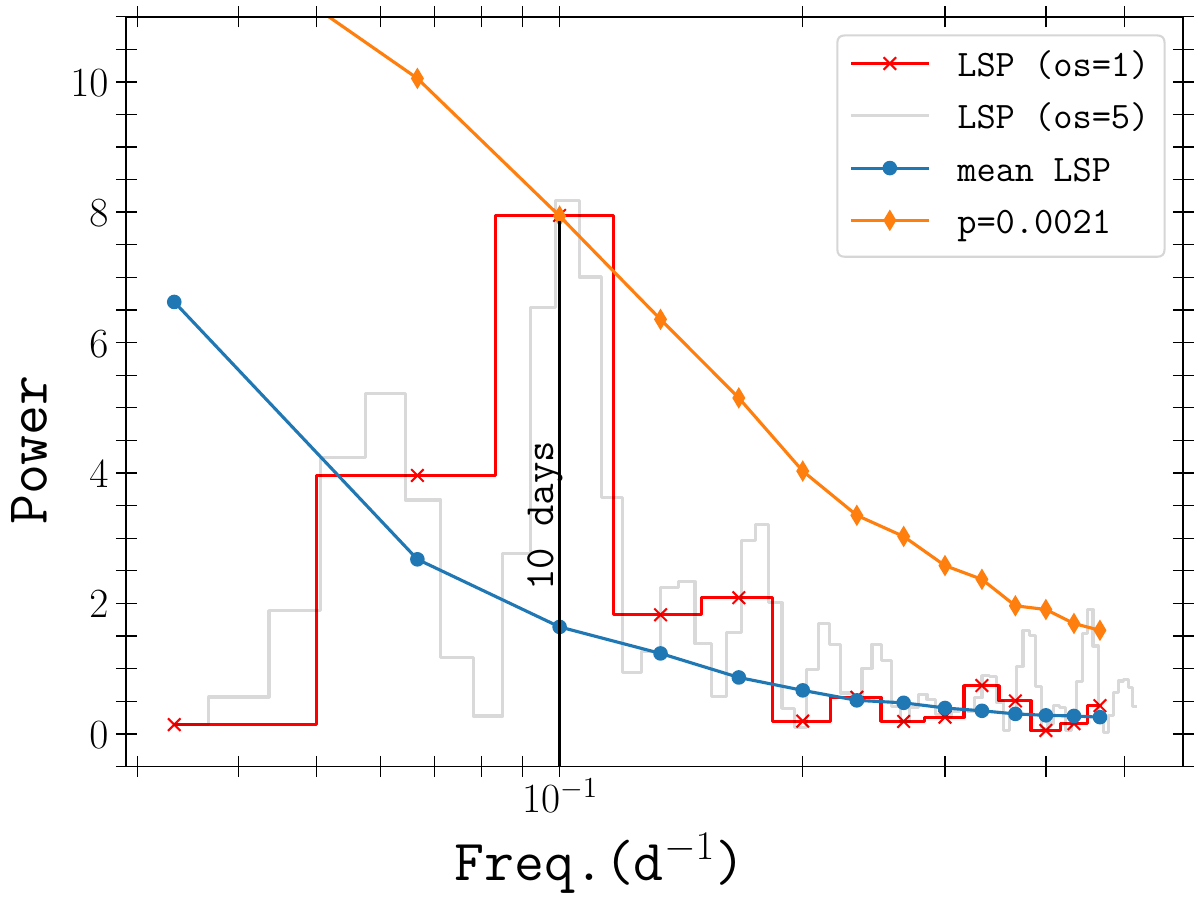}
    \includegraphics[width=0.9\linewidth]{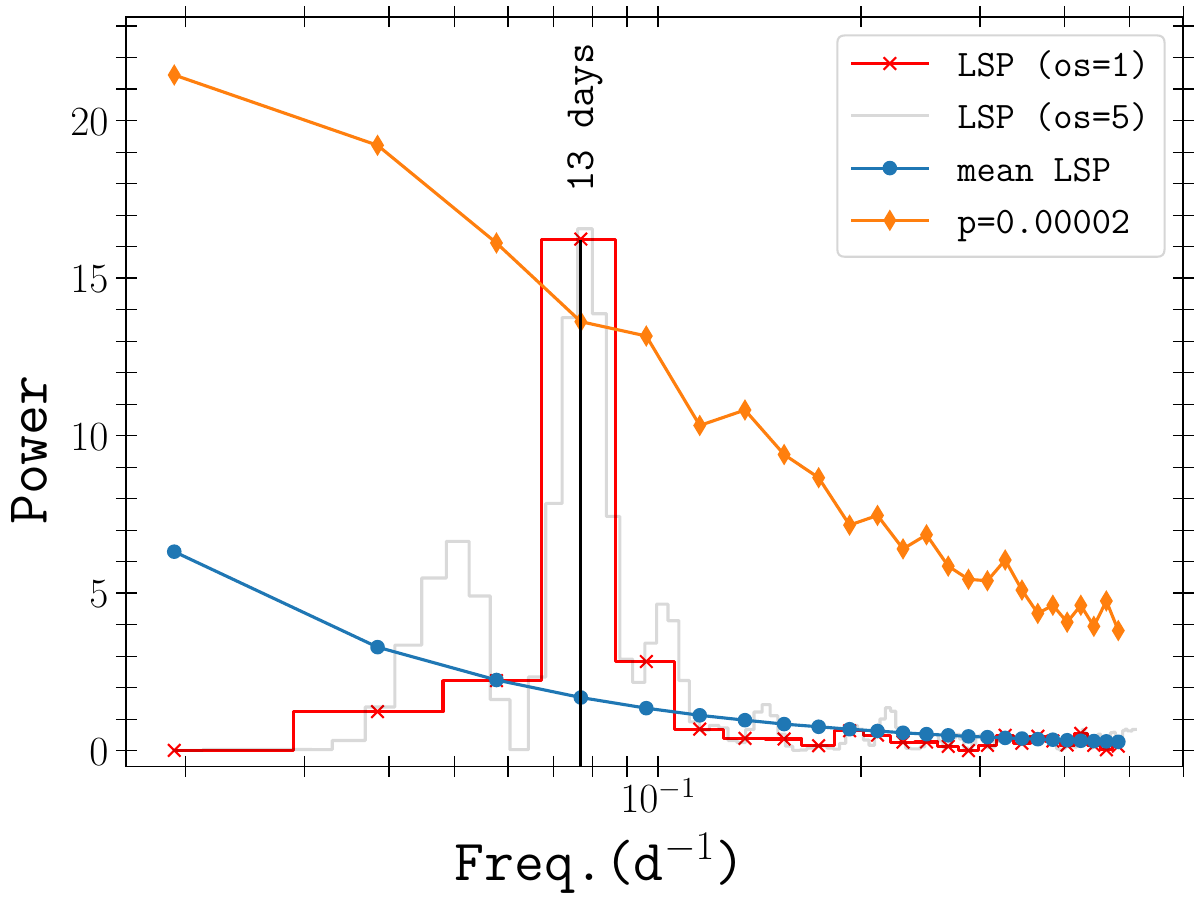}
    \caption{LSP for the $\gamma$-ray light curve for the epoch MJD~61167 -- 61200 (top) and that of the detrended light curve for the epoch MJD~61205 -- 61257 (bottom). `os' represents the oversampling factor.}
    \label{fig_3}
\end{figure}
\begin{figure}
    \centering
    \includegraphics[width=0.9\linewidth, trim=0 1cm 0 0, clip]{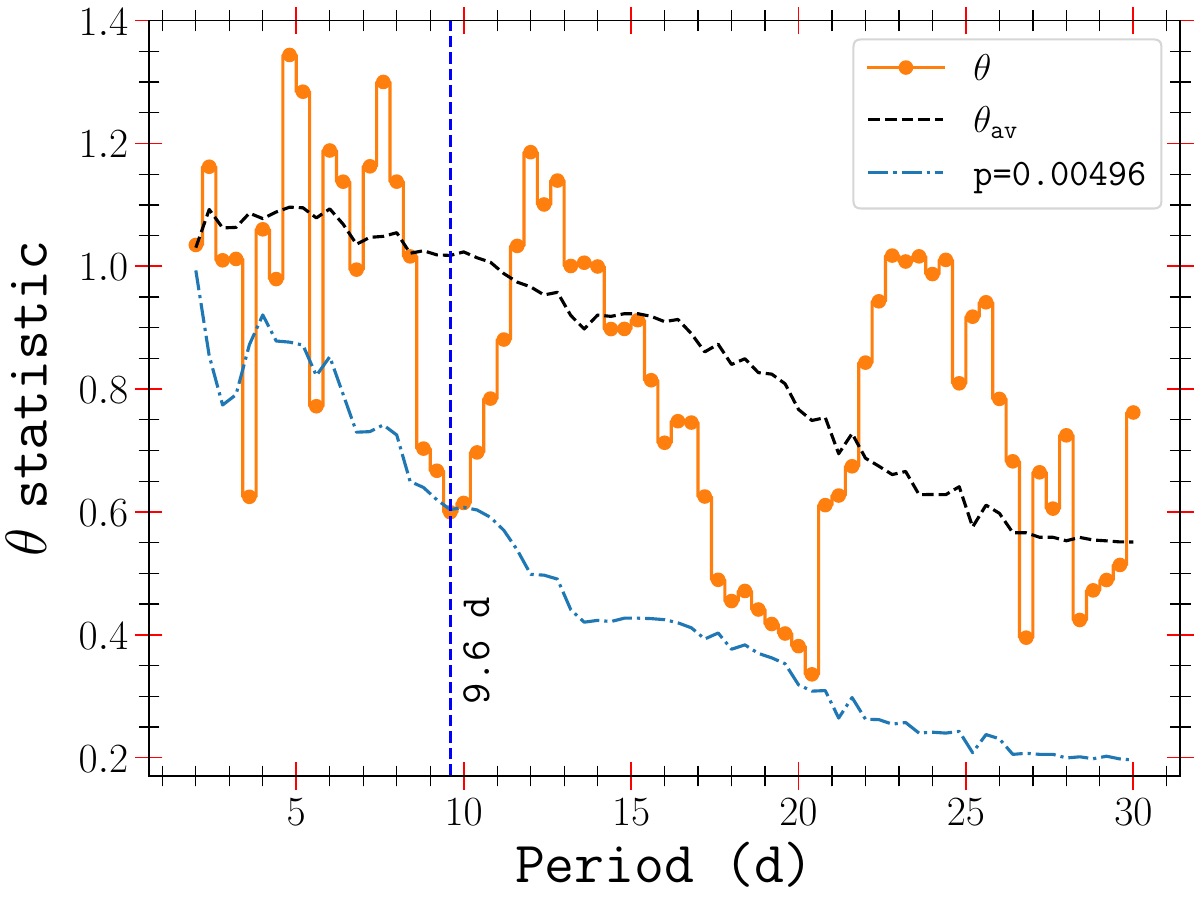}
    \includegraphics[width=0.9\linewidth]{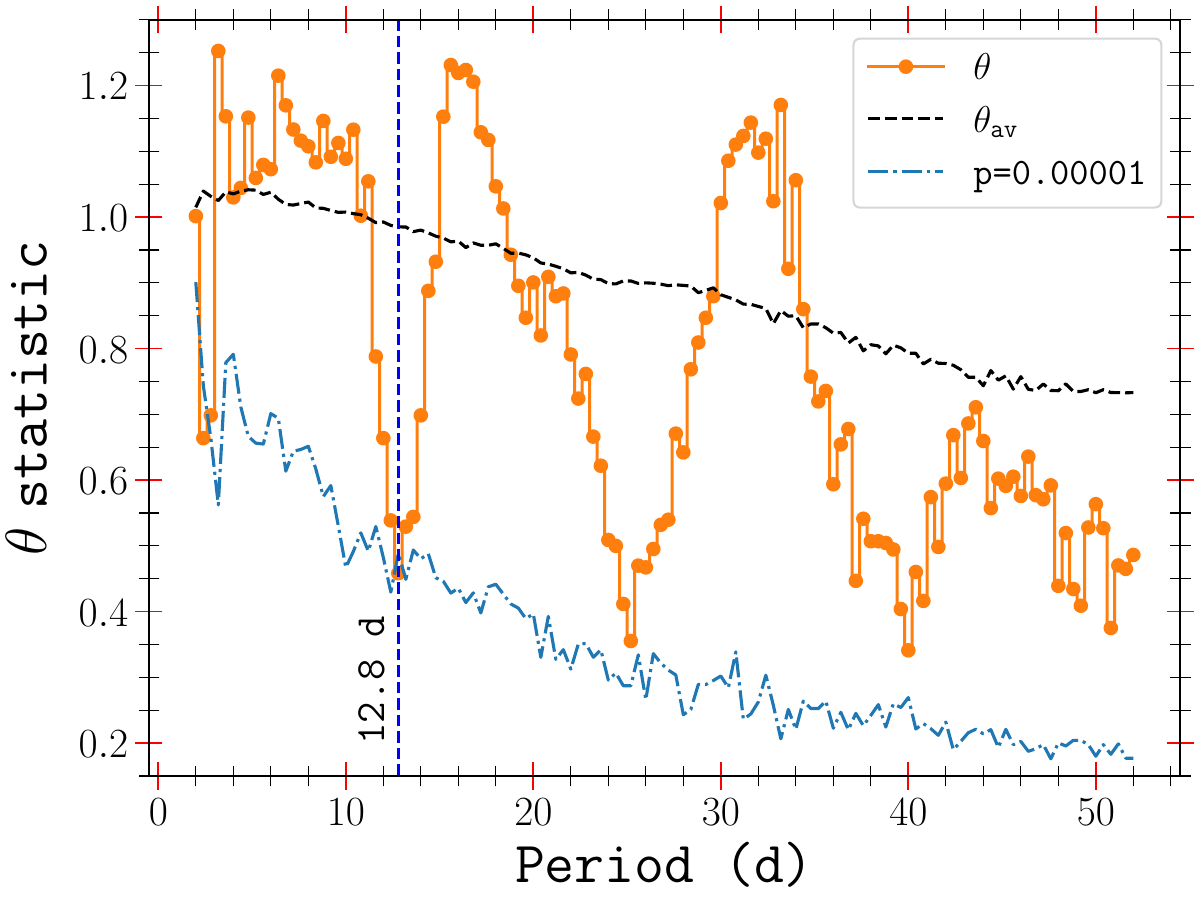}
    \caption{PDM for the $\gamma$-ray light curve for the epoch MJD~61167 -- 61200 (top) and that of the detrended light curve for the epoch MJD~61205 -- 61257 (bottom).}
    \label{fig_4}
\end{figure}
\begin{figure}
    \centering
    \includegraphics[width=0.9\linewidth]{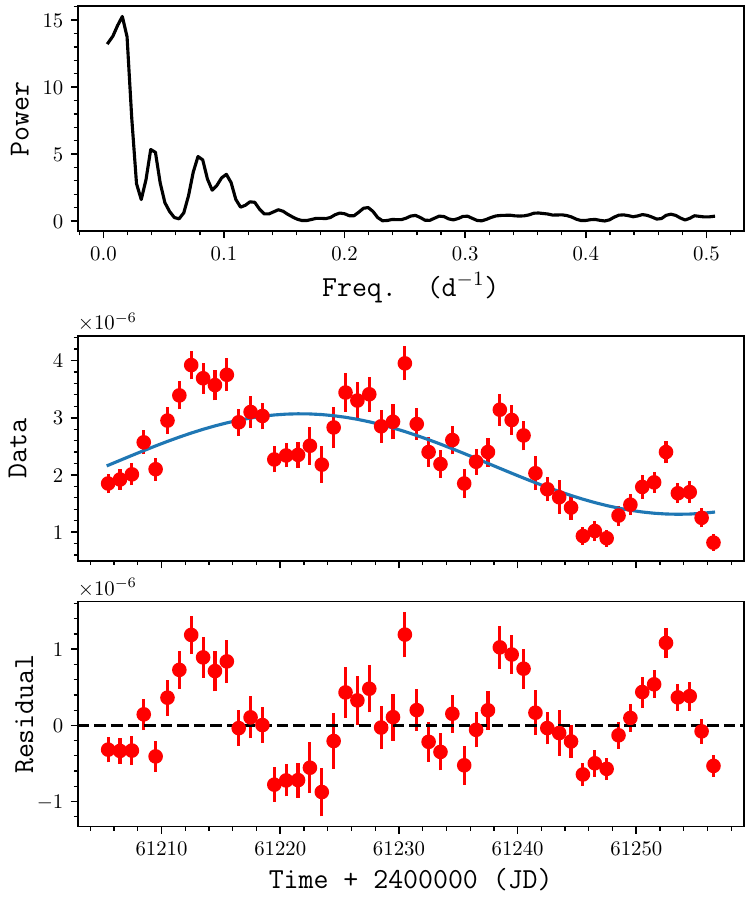}
    \caption{Top: LSP of light curve for the epoch MJD~61205 -- 61257, showing very high power at the lower frequency end of the periodogram caused by the long-term rising and decaying trends of the flare. An oversampling by a factor of five in the frequency domain was employed to get a more precise value of this frequency. Middle: $\gamma$-ray data along with the long-term trend associated with high power at the lower frequency end within the noted epochs. Bottom: Residual (Detrended) light curve over which LSP is again computed for significance estimation.}
    \label{fig_5}
\end{figure} 
\begin{figure}
    \centering
    \includegraphics[width=0.9\linewidth, trim = 0 0 0 1.4cm, clip]{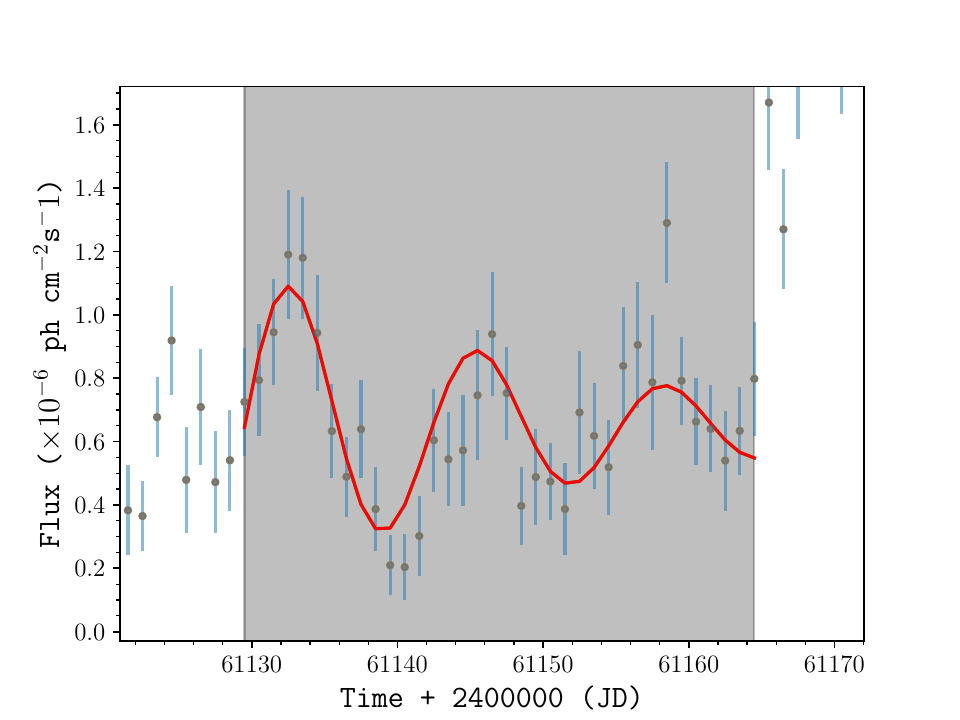}
    \caption{Light curve fitting using the damped oscillatory function (eq.~\ref{dp_os}) for the epoch  MJD~61129 -- 61165.}
    \label{fig_6}
\end{figure}

\noindent
We mainly constrained our interest in the region, R2 (see Fig.~\ref{fig_1}), which covers the flare region shown by the object. The flare starts at around  MJD~61167 and appears to be composed of two distinct components, where the first one ends by MJD~61200, followed by a recommencement of another component that lasts at around MJD 61257. 
We first implemented the WWZ analysis for the complete LC to get a general overview of frequency powers at all the epochs, which further assisted with segmenting the epoch spans over which LSP analysis could be employed to get the maximum contribution of the periodic signals in the LC. Fig.~\ref{fig_2} represents the two-dimensional WWZ map of the complete  LC, which assertively reveals the existence of three prominent powers near frequency 0.1~d$^{-1}$, maximizing at around epochs MJD~61137, 61186, and 61237. The first one belongs to R1, while the latter two belong to R2. A secondary high power appears near a lower frequency 0.05~d$^{-1}$ at around MJD~61202, which is highly likely a chance coincidence due to the fact that it maximizes precisely near the end of the first component of the flare and beginning of the second component, along with the certain prominent frequency feature at around 0.1~d$^{-1}$ (hence acting like a lower order harmonic of 0.1~d$^{-1}$). The distinct shapes of the decaying and rising parts of the two components of the flare may also give rise to high power at 0.05~d$^{-1}$. Consequently, we separately tested the first three prominent features only with other complementary methods (LSP and PDM). It is worth noting that the two features under consideration in R2 appear to have mild power. This is mainly because the existence of other features (such as 0.05~d$^{-1}$) at simultaneous epochs can drastically diminish normalized powers in the WWZ map.
After the basic scrutiny of the WWZ map, we used LSP (a sinusoidal-based approach) and PDM (a non-sinusoidal approach) to confirm the preliminary results with the WWZ. The significances of the periodicities deduced with these complementary methods were estimated by generating 50,000 artificial  LCs having probability distribution function (PDF) and power spectral density (PSD) slopes similar to segmented  LCs containing the periodicity features following \cite{2013MNRAS.433..907E}, and evaluating the corresponding percentile ($p$) value of the signals. We note here that while estimating the underlying PSD shape, since high powers at peculiar QPO frequencies adversely affect the PSD fitting parameters (see \cite{2026MNRAS.550g1284K}), true shapes could be estimated at the expense of the removal of that point. Thus, we removed the individual QPO frequency point for all three PSD shape inspections. With the above as the general approach, the tweaks, any additional or alternative approach (as done here for R1), and the results of our analyses for these three epochs (one as R1 and two in R2) are as follows. 
\noindent
\paragraph {Findings for MJD~61167 -- 61200~}
 ~In the LSP analysis, without any oversampling, we find a high power peaking at around a period of 10~d. To further improve the precision of this period, we also implemented an oversampling of frequency by five times and found it to be around 9.8~d. Following the simulation process ($5\times10^4$ runs), the frequency peak was found to have a significance $p$-value of $\sim$0.0021 (see the top panel of Fig.~\ref{fig_3}), which is equivalent to $\sim$3.075$\sigma$. Analysis with the PDM of this segment, a broad and significant dip was found at around 9.6~d, having a $p$-value of 0.00496 (see the top panel of Fig.~\ref{fig_4}). The PDM shows another such dip at around 19~d with a wider spread, appearing to be a harmonic of the first one, though with a little lower significance. There are some highly significant dips visible at lower timescales, but they could not be confirmed with LSP and are likely noise. 
 \paragraph{Findings for MJD~61205 -- 61257~}~
 For this epoch span, a long-term trend is clearly visible in the  LC. In the evaluated LSP, this appeared as a very high power peak at the lower frequency region in its periodogram (see the top panel of Fig.~\ref{fig_5} ). Therefore, we first detrended the LC to get rid of this unwanted signal, a process which is generally termed as prewhitening. Fig.~\ref{fig_5} includes the LSP of this portion of the LC, showing high power at the lower frequency end, along with the trend and the detrended LC (see the middle panel of Fig.~\ref{fig_5} ). With the detrended (or the residual)  LC (see the bottom panel of Fig.~\ref{fig_5}), we again computed its LSP to find the QPO period and its significance following the simulation process similar to the previous one, and found a strong periodic signal of 13~d (see the bottom panel of Fig.~\ref{fig_3}) without oversampling. An oversampling by a factor of five gave a more precise value of $\sim$12.8~d. With a simulation run of 50,000, this signal was found to have an at least significance $p$-value of $<2\times10^{-5}$, equivalent to $>4.26\sigma$. The PDM analysis of the detrended LC also gave a significant dip at $\sim$12.8~d with $p$-value of $10^{-5}$ (see the bottom panel of Fig.~\ref{fig_4}), along with two other similar significant dips at around 25~d and 40~d, highly likely as harmonics of the first one.
 \paragraph{Findings for MJD~61129 -- 61165~}~
 Visual inspection of the  LC within these epochs reveals an oscillatory feature associated with damping, with a period of the order of 10~d. However, we could only find a small elevation of power in its LSP. This could be mainly because of the relatively low amplitude of oscillation, the dominance of uncertainty noise over the flux, or the fact that there are not enough significant periodic cycles due to damping. So, we employed a slightly different approach to assess its periodic nature. First, we directly fit this epoch span (see Fig.~\ref{fig_6}) with a damped sinusoidal function given as 
 \begin{equation}
 \label{dp_os}
     y(t)=Ae^{-t/\tau}sin(\frac{2\pi t}{P})+c~,
 \end{equation}
 where $y$ is the flux at epoch $t$, with $A,~ \tau,~P,~\text{and}~c$ being normalization, decay time scale (maybe related to some coherence length of the emitted plasma), period of oscillation, and mean flux level, respectively. The parameters of interest here are $P~\text{and}~\tau$, which were found to be $12.9\pm0.3~\text{d and } 21.2\pm8.6~\text{d}$. The other parameters $A~\text{and}~c$ were $(5.2\pm1.2)\times10^{-7}~\text{and }(6.4\pm0.3)\times10^{-7}$. For estimating the significance of this damped oscillatory feature, we used the traditional $\chi^2$ statistic, given as 
 \begin{equation}
     \chi^2=\sum\frac{(y_i-\bar{y})^2}{2\sigma^2_i}
 \end{equation}
 where $\bar{y}$ is given by eq.~\ref{dp_os} corresponding to the derived fitting parameters, and $\sigma_i$ is the uncertainty associated with the flux point $y_i$. Similar to those with LSPs, we generated 50000 artificial LCs based on the PDF and PSD of the  LC within these epochs and computed the $\chi^2$-value of each of those with respect to $\bar{y}$ to get a distribution of 50000 $\chi^2$. This way, the $\chi^2$ of the original  LC within the noted epochs was found to have a $p$-value of $\sim$$3\times10^{-4}$.  
 
 \section{Discussion AND CONCLUSIONS}
 \label{section5}
\noindent
In this paper, we investigated the existence of $\gamma$-ray QPOs in the  CL blazar B2 1420$+$326 using three well-suited techniques: WWZ, LSP, and PDM. Epoch selection of $\gamma$-ray data was made from January 1, 2026, to August 25, 2026, during which the source passed through quiescent, moderately, and highly variable phases. Particularly, the source entered a strong flaring phase post MJD~$\sim$61167 (May 7, 2026) lasting for around two and a half months. We found potential QPO features of a period of the order of 10~d during the flare and a mildly oscillatory behavior of similar timescale associated with a damping (timescale of order of 20~d) in the relatively quiescent phase before the flare. The damping timescale is highly uncertain and could be due to factors including low oscillation amplitudes, fewer oscillation cycles, relatively high flux uncertainty compared to the flux values themselves, and, obviously, visual epoch selection bias. Though it can't be uniquely constrained, this timescale may be associated with a coherence length after which the plasma responsible for emission gets randomly diffused. \\
\\
The observed emissions in blazars are typically dominated by the relativistic jet owing to high Doppler boosting, often exceeding the thermal contribution from the accretion disc. Possible jet based models responsible for a $\gamma-$ray flare include shock-in-jet scenario \citep{1985ApJ...298..114M} which are caused by variations in the plasma injection rate, magnetic fields and bulk Lorentz factor of the flow, relativistic magnetic reconnections leading to very fast emission changes \citep{2020NatCo..11.4176S, 2022MNRAS.510.3641R}, and mini-jet or jet-in-jet, where highly compact regions with very high Doppler factors are responsible for substantial flare \citep{2009MNRAS.395L..29G}. In the previous studies, several QPO models have been suggested, including accretion-disc based origins such as disc oscillations, hot spots, and instabilities \citep{Mangalam1993, Chakrabarti1993, Espaillat2008, 2009ApJ...690..216G, Kishore_2023}, and Lense–Thirring precession of the inner accretion disc \citep{1998ApJ...492L..59S, 2018MNRAS.474L..81L}. The accepted jet-based QPO origins include helical motion of the emitting plasma \citep{2018NatCo...9.4599Z, 2021MNRAS.501...50S}, jet precession \citep{Rieger2004, Caproni_2017}, and magnetohydrodynamic instabilities \citep{Li_2004, Arras_2006, 2020MNRAS.494.1817D, 2022Natur.609..265J, h55b-mdp5}. Among these, jet precession, precession of the inner disc in a binary supermassive black hole system, or Lense–Thirring precession of the inner accretion disc, lead to long-term QPOs. Short-lived or transient QPOs in the $\gamma$-ray, however, could appear only due to the processes taking place inside the jet. \\
 \\
The similar observed periods of QPOs in three nearby epochs of the LC during MJD 61129 -- 61257 suggest a multi-component emission possibly governed by the same physical scenario, differing likely only in the amplitude of flux variation.  Detection of a QPO of order $\sim$10~d significantly during flare epochs rules out persistent QPO models in this scenario.
%
One possible explanation could be a helical jet geometry framework \citep{ Mohan_2015, 2017MNRAS.465..161S,2018NatCo...9.4599Z,2021MNRAS.501...50S,2022MNRAS.510.3641R}, where an emitting plasma blob is moving in a helical trajectory along the jet. Differences in the axial distance of such blobs from the jet spine will result in minor variations in the gyration period. The two components in the flare showing distinct QPO periods here may correspond to two separate plasma blobs injected with different axial separations or pitch angles. Such motion of the blob results in a periodic change in its viewing angle ($\theta_{obs}$:  angle of the blob velocity with the observer's line of sight), given as 
\[
\cos \theta_{\rm obs}(t)
=
\sin\phi \sin\psi
\cos\left({2\pi t}/{P_{\rm obs}}\right)
+
\cos\phi \cos\psi~,
\]
 where $t, ~\phi, ~\psi, ~P_{\rm obs}$ are time, pitch angle of the helix, angle between the jet axis and the observer's line of sight, and observed QPO period from the  LC, respectively \citep{2018NatCo...9.4599Z}.
Variation in $\theta_{obs}$ is reflected in the Doppler factor
($\delta = 1/[\Gamma(1-\beta\cos\theta_{\rm obs})]$,
where $\Gamma=(1-\beta^2)^{-1/2}$ is the bulk Lorentz factor and $\beta$ is the velocity),
which may lead to QPOs with periods ranging from a few days to months. The physical period $P$ in the blob rest frame is given by $P_{\rm obs}$ as
$P = P_{\rm obs}/(1-\beta\cos\phi\cos\psi)$. The main drawback of this model is that it can account for a flux oscillation of invariable amplitude, contrary to the signatures observed during the period MJD 61167 -- 61257. \cite{2021MNRAS.501...50S} explained the nearly 47-day $\gamma$-ray QPO with changing amplitude in 3C 454.3 using a curved-jet framework in which the blob moves within a curved helical jet rather than a straight one, making a time-dependent angle between the jet axis and the observer's line of sight ($\psi=\psi(t)$). Using $P_{obs}=$10 d obtained from our analysis, $\Gamma=$ 19, $\psi=1.1^{\circ}$ from \cite{2021A&A...647A.163M} and assume $\phi=1^{\circ}$ from \citep{2018NatCo...9.4599Z}, we estimated physical period $P=15.9$ years in the blob rest frame. During this period, the distance traveled by the blob is $D=c\beta\;P\,cos(\phi)=$ 4.9 pc per cycle; the total distance over the 7 cycles of QPOs (see Fig.\,\ref{fig_1}) during the period MJD 61167-61257 is 34.3 pc, and the total projected distance over 7 cycle would be $D_p = 7D\sin\psi=0.65$ pc=0.09 mas.  Earlier VLBA observations of B2 1420+326 revealed a parsec-scale jet containing the radio knot K20, with an average size of \(0.084 \pm 0.015\) mas \citep{2021A&A...647A.163M} which is comparable with the moving emitting plasma blob interpretation. However, the available VLBA observations are not enough to reveal the helical-curved geometry of the jet. The helical-curved jet model could be a possible interpretation of the 10-day-long $\gamma$-ray QPO in B2 1420+326. The LC of the source during MJD 61167--61257, shown in Fig. \ref{fig_1}, exhibits a gradual increase over the first five cycles, followed by a rapid damping during the sixth and seventh cycles. This type of variability may require a relatively low curvature of the jet, and similar evidence has been reported in \citet{2022MNRAS.510.3641R}. 
\\\\
Another possible scenario for producing short-lived quasi-periodic variability in relativistic jets is magnetic reconnection within a sequence of nearly equispaced magnetic islands (plasmoids) \citep{2013RAA....13..705H, PhysRevLett.132.035101}. As explained by \cite{2020NatCo..11.4176S}, current-driven kink instabilities can distort the ordered structure of the jet and produce turbulence and filamentary current sheets.  These current sheets may subsequently become unstable, undergo magnetic reconnection, and fragment into multiple plasmoids. These multiple plasmoids can produce repeated acceleration of particles. Hence, we may get quasi-periodic variability.
 A $\gamma$-ray QPO on a time scale of 3.6 d in PKS 1510$-$089 has been explained by \cite{2022MNRAS.510.3641R} with the periodic occurrence of reconnection events triggered by quasi-equidistant magnetic islands. Simulations of relativistic reconnection suggest that plasmoids can produce flares lasting for hours or even days, and maintain active behavior on several days or longer \citep{Giannios_2013,Petropoulou_2016}. 
 This model appears to explain nearly 10-day transient QPOs with the reconnection/plasmoid hypothesis. \\
\\
One of the most plausible explanations for transient QPOs is also suggested by \cite{2022Natur.609..265J} with kink instabilities (KI) based on the method given by \cite{2020MNRAS.494.1817D}.  
These kinds of instabilities can develop in a jet with a strong toroidal magnetic field and cause both twisting of field lines and an increase in particle acceleration \citep{2009ApJ...700..684M,2017MNRAS.469.4957B, PhysRevLett.121.245101,2024MNRAS.527.9132T}.  Quasi-periodicity on timescales of hours to days can be given by the evolution of a current-driven KI within a magnetically dominated relativistic jet. The kink growth time period is given by \cite{2020MNRAS.494.1817D}  as 
$T_{\rm obs} = {R_{\rm KI}}\,/{v_{\rm tr}\,\delta}
$ where ${R_{\rm KI}}, {v_{\rm tr}}$ are the emission-region size and average propagation jet velocity, respectively. Using the $\delta=33$  and $\delta=20$ from \cite{2021A&A...647A.163M} and \citep{2021MNRAS.502.5245P}, respectively, a typical emission-region size (${R_{\rm KI}}$)  of $10^{16}-10^{17}$ cm, and $v_{\rm tr}=0.16c$ from \cite{2020MNRAS.494.1817D}, the corresponding values of $T_{\rm obs}$  are  0.73-7.3 d and 1.21-12.1 d, respectively. The time period of the observed QPOs is consistent with the evolutionary timescale of KI. \\
\\
We investigated various AGN emission models and found that a few of them, e.g., helical jet geometry, magnetic reconnection, and kink instabilities, could explain our findings.
Narrowing down to further constrain the underlying process would certainly benefit from a multiwavelength approach. Particularly, polarization inspections to trace the possible magnetic field variations and radio maps assessments for comprehending the ejection of any new plasma components, along with their evolution. Since  CL blazars are rare, and the detection of QPOs in these is even rarer, the search for QPOs in different EM band LCs of CL blazars can emerge as a new key research field.     
\\\\
{\it Acknowledgment:} This work made use of data from the Fermi-LAT available in the LAT data server at: \url{https://fermi.gsfc.nasa.gov/ssc/data/access/} \\
\\
{\it Software:} lightkurve \citep{2018ascl.soft12013L}, 
SciPy \citep{2020SciPy-NMeth}, PyAstronomy \citep{2019ascl.soft06010C}. The Fermi-LAT data analysis software is available at \url{https://fermi.gsfc.nasa.gov/ssc/data/analysis/software/}


\bibliography{References}

\bibliographystyle{aasjournalv7}
\end{document}